\documentclass[runningheads]{llncs}
\usepackage[T1]{fontenc}
\usepackage{graphicx}
\usepackage{xcolor}
\usepackage{orcidlink}
\begin{document}
\title{ASPIRE: Assistive System for Performance Evaluation in IR}
\titlerunning{ASPIRE: Assistive System for Performance Evaluation in IR}
\author{Georgios Peikos\inst{1}\orcidlink{0000-0002-2862-8209}\and
Wojciech Kusa\inst{2}\orcidlink{0000-0003-4420-4147}\and
Symeon Symeonidis\inst{3}\orcidlink{0000-0002-3259-614X}}
\authorrunning{Peikos et al.}
\institute{University of Milano-Bicocca, Milan, Italy \\
\email{georgios.peikos@unimib.it} \\
\and
Allegro, Warsaw, Poland \\
\email{wojciech.kusa@allegro.com}\\
\and
Democritus University of Thrace, Department of Production and Management Engineering, Xanthi,Greece\\
\email{symesyme@pme.duth.gr}
}
\maketitle
\begin{abstract}
Information Retrieval (IR) evaluation involves far more complexity than merely presenting performance measures in a table.
Researchers often need to compare multiple models across various dimensions, such as the Precision-Recall trade-off and response time, to understand the reasons behind the varying performance of specific queries for different models.
We introduce ASPIRE ({Assistive System for Performance Evaluation in IR}), a visual analytics tool designed to address these complexities by providing an extensive and user-friendly interface for in-depth analysis of IR experiments. 
ASPIRE supports four key aspects of IR experiment evaluation and analysis: single/multi-experiment comparisons, query-level analysis, query characteristics-performance interplay, and collection-based retrieval analysis.
We showcase the functionality of ASPIRE using the TREC Clinical Trials collection.
ASPIRE is an open-source toolkit available online\footnote{\url{https://github.com/GiorgosPeikos/ASPIRE}}.

\end{abstract}
\section{Introduction}
Evaluating the effectiveness of Information Retrieval (IR) systems is a core aspect of IR research, generally conducted based on the Cranfield paradigm for offline experimental evaluation~\cite{cleverdon1962aslib,voorhees2019evolution}.
This evaluation requires three essential components: (1) a collection of information items, e.g. \textit{documents}, (2) \textit{queries} representing information needs, and (3) relevance judgments (also called \textit{qrels}), specifying the relevant items for each query.
Given document rankings produced by IR systems, the components mentioned above enable the estimation of quantitative performance measures to determine which systems are more effective, either overall or for specific queries.

To evaluate IR systems, researchers have proposed a wide range of performance measures~\cite{jarvelin2017ir}, along with several tools to assist in their calculation, such as the $trec\_eval$ tool\footnote{\url{https://github.com/usnistgov/trec\_eval}}, among others~\cite{bassani2022ranx,macavaney2022streamlining,palotti2019trectools,10.1145/3331184.3331398,DBLP:conf/sigir/FrobeRMDRB0HP23}.
Nonetheless, assessing a system's performance extends beyond examining average and per-query effectiveness measures. It often necessitates a detailed, fine-grained analysis, such as analyzing performance across queries of varying complexity or based on groups of contextually similar queries~\cite{voorhees2001philosophy}.

Visual analytics can aid researchers conduct these in-depth analyses of IR system behavior~\cite{ferro2019visual}, leading to the development of numerous tools over the years~\cite{DBLP:conf/sigir/ArmstrongMWZ09a,DBLP:conf/cikm/DittenbachPPRB09,DBLP:conf/ictir/Yang016,DBLP:conf/sigir/AmigoAAGRV17,DBLP:journals/tmm/IoannakisKPC18,DBLP:journals/ipm/AngeliniFFSS18,DBLP:conf/ecir/HofstatterZH20,tamannaee2020vis,DBLP:conf/sigir/JoseNM0Y21,gonzalez2023exploratory}.
Some of these tools are no longer available online~\cite{DBLP:conf/sigir/ArmstrongMWZ09a,DBLP:conf/cikm/DittenbachPPRB09,DBLP:conf/ictir/Yang016,DBLP:conf/sigir/AmigoAAGRV17}, while others are tailored to specific evaluation analyses, such as query/document content-based analysis~\cite{DBLP:conf/ecir/HofstatterZH20}, analysis of queries where systems' retrieval behavior diverges~\cite{DBLP:conf/sigir/JoseNM0Y21}, round-based retrieval evaluations, focused on tracking differences in systems' performance across evaluation rounds~\cite{gonzalez2023exploratory}, or identification of preferred IR system configurations~\cite{DBLP:journals/ipm/AngeliniFFSS18}.
RETRIEVAL is an online platform for evaluating IR methods based on the Cranfield paradigm, developed in 2018~\cite{DBLP:journals/tmm/IoannakisKPC18}. 
The tool allows users to select performance evaluation measures and offers downloadable reports of these evaluations.
Similarly, Vis-Trec, an open-source tool designed for offline retrieval evaluation, offers a range of valuable features, including diverse performance visualizations, query difficulty analysis, and result reporting to LaTeX-compatible tabular formats~\cite{tamannaee2020vis}.
However, the systems' capabilities could be improved, for instance, by supporting a complete range of evaluation measures, analyses, and by leveraging web-based interactive graph creation techniques to aid users in evaluating retrieval behavior.

This paper introduces ASPIRE (Assistive System for Performance Evaluation in IR), a web-based visual analytics tool with an extensive, user-friendly interface for in-depth IR experiment analysis.
ASPIRE currently supports four key aspects of system evaluation: single-experiment evaluation, multiple-experiment comparisons, and query-based and collection-based analyses.
It is designed to help researchers gain deeper insights into IR system performance and simplify evaluation tasks.
ASPIRE is available online\footnote{\url{https://aspire-ir-eval.streamlit.app/}}.
For scenarios involving private data, users can download and operate ASPIRE locally\footnote{\url{https://github.com/GiorgosPeikos/ASPIRE}}.
For a quick overview of its features, please refer to the demo videos\footnote{\url{https://www.youtube.com/playlist?list=PLyHXxSV5Xhvhap75xMfXCaV1MQsJD0gAz}}.

\section{ASPIRE: Implementation Details and Overview}
\label{sec:aspire_overview}
\begin{figure}[t]
    \centering
    \includegraphics[width=\linewidth]{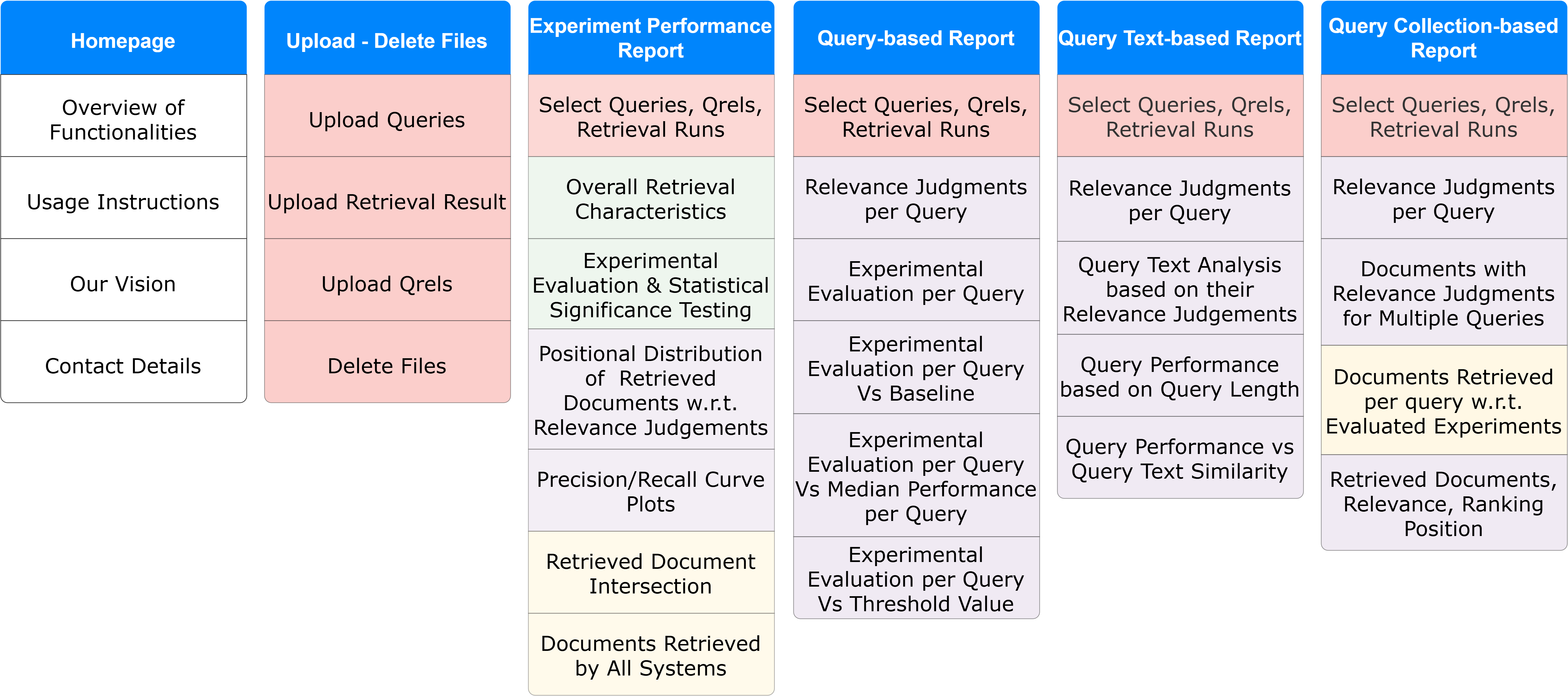}
    \caption{ASPIRE's functionalities. Each column is a different web page, and each block is a section of a web page. Red blocks indicate user actions, green blocks represent performance evaluation result tables, purple blocks represent plots and analysis, and yellow blocks show analysis.}
    \label{fig:overview}
\end{figure}
ASPIRE is developed in Python, with \textit{streamlit (v1.37.0)} powering its interactive web interface.
Its high-level architecture follows a modular structure, with distinct components organized across multiple web pages, each divided into clearly defined sections (see Figure~\ref{fig:overview} for overview).
This design enables extensibility, allowing the integration of new analyses by adding sections within existing pages or creating new pages, ensuring adaptability and compatibility with future IR evaluation practices.

ASPIRE uses \textit{ir\_measures} to evaluate IR experiments and employs \textit{statistics} and \textit{statsmodels} for statistical analysis of experiment results.
For visualization, \textit{plotly-express} enables fast, interactive plotting features, while the \textit{transformers} library is used for query performance evaluation w.r.t. to contextually similar queries.
To use ASPIRE, users can upload standard TREC-style files, i.e. a query file (i.e. qid and query text), a qrels file (with qid, iteration (Q0), doc\_id, and relevance judgments), and many TREC-style runs (with qid, Q0, doc\_id, rank, score, and experiment\_id).
Each input is checked for missing values, format inconsistencies, and column name variations.
Outputs include exportable plots in PNG format, tables in CSV format, and the complete analysis summary as a PDF.
As shown in Figure~\ref{fig:overview}, ASPIRE currently features two configuration web pages and four pages for experimental result analyses, each dedicated to a unique aspect of IR experiment evaluation.
On each page, users select their uploaded TREC-style files and initiate their analysis by pressing a button.
During the analysis, they can adjust parameters, like measures or baseline runs, with real-time updates to visualizations and tables.
The \textbf{Experiment Performance Report} assesses IR experiment outcomes with standard measures, supporting multi-run comparisons, statistical significance testing, and precision-recall curve generation.
The \textbf{Query-based Report} offers query-level performance insights, while the \textbf{Query Text-based Report} examines the link between query characteristics (e.g., length) and effectiveness, providing word clouds and 2D/3D query similarity visualizations.
The \textbf{Query Collection-based Report} analyzes relevance judgments distribution, identifying documents relevant across multiple queries and visualizing document rankings.
Two additional evaluation aspects are under development, focusing on query performance prediction~\cite{meng2023query} and multidimensional relevance~\cite{peikos2024systematic}.
\begin{figure}[h]
    \centering
    \includegraphics[width=0.95\linewidth]{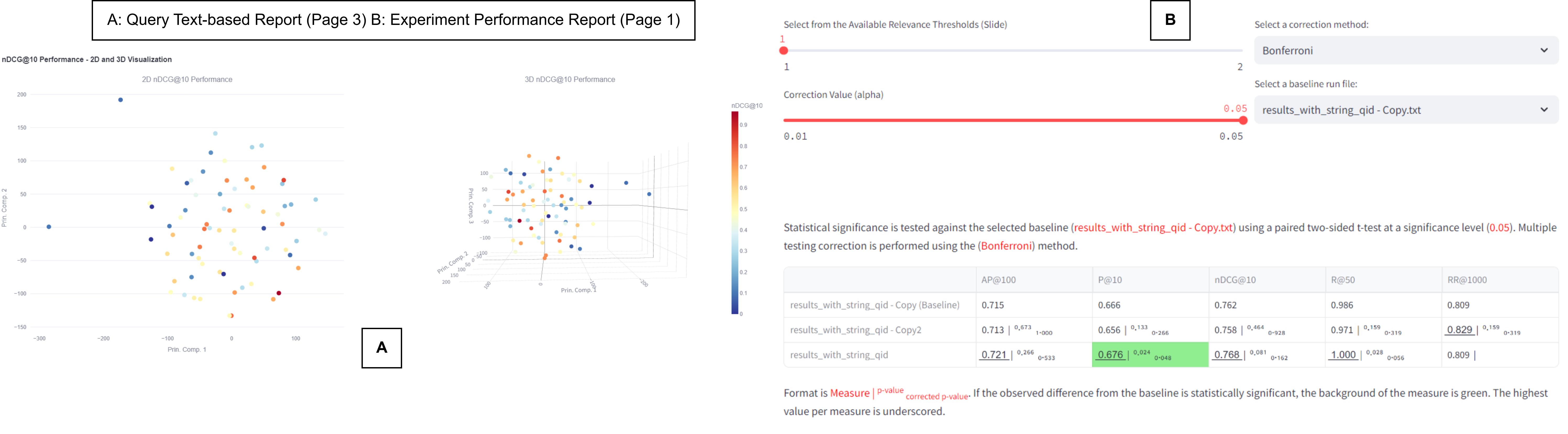}
    \caption{ASPIRE's user interface and examples of its functionality using runs submitted in TREC Clinical Trials 2021~\cite{peikos2022unimib}.}
    \label{fig:single-experiment-ui}
\end{figure}
ASPIRE is built to handle large collections and evaluate numerous runs simultaneously, accommodating, for example, all participant submissions in shared tasks of evaluation initiatives.
\section{Use Cases}
ASPIRE is designed to support various users in the IR community, providing streamlined analysis capabilities for researchers, shared-task organizers, and practitioners.
Researchers can easily upload their files—query, qrels, and run files—to obtain immediate results, interactive visualizations, and downloadable insights, enabling efficient analysis of their IR experiments.
ASPIRE complements existing IR run repositories~\cite{DBLP:conf/sigir/0002VS24,bassani2023ranxhub} that promote reproducibility by allowing users to retrieve stored runs, import them, and conduct in-depth comparisons with their experiments.
For shared task organizers, ASPIRE facilitates single-run and multi-run evaluations, with plans to introduce a dedicated page in future versions to focus explicitly on their use-cases.
Figure~\ref{fig:single-experiment-ui} showcases some of the functionalities currently integrated in ASPIRE.
\section{Conclusions and Future Work}
ASPIRE addresses a relevant need for visual interpretation and understanding of IR system performance.
This need has been increasingly recognized as a valuable goal within the IR community, similar to the growing emphasis on explainability and interpretability of IR systems.
Our tool provides a modular design that enables easy implementation and integration of IR evaluation practices, facilitating their broader adoption within the IR community.
By leveraging shared retrieval experiment data alongside publications (i.e. run files, qrels, queries), ASPIRE enables researchers and readers to explore and analyze experiment results beyond reported performance scores, fostering a deeper engagement with the data.
In the future, ASPIRE could play an essential role in encouraging the adoption of good practices in IR evaluation, driving forward the community’s commitment to transparency and reproducibility.
\begin{credits}
\subsubsection{\ackname}
This work was funded by the National Plan for NRRP Complementary Investments (PNC, established with the decree-law 6 May 2021, n. 59, converted by law n. 101 of 2021) in the call for the funding of research initiatives for technologies and innovative trajectories in the health and care sectors (Directorial Decree n. 931 of 06-06-2022) - project n. PNC0000003 - AdvaNced Technologies for Human-centrEd Medicine (project acronym: ANTHEM). This work reflects only the authors’ views and opinions, neither the Ministry for University and Research nor the European Commission can be considered responsible for them.
\subsubsection{\discintname}
The authors have no competing interests to declare that are relevant to the content of this article.
\end{credits}

\bibliographystyle{splncs04}
\bibliography{ref}
\end{document}